\documentstyle[epsfig,12pt]{article}
\def\v1{\vspace{1cm}}
\def\be{\begin{equation}}
\def\ee{\end{equation}}
\def\bc{\begin{center}}
\def\ec{\end{center}}

\def\vh{\varphi}

\newcommand{\bea}{\begin{eqnarray}}
\newcommand{\eea}{\end{eqnarray}}

\begin{document}
\setlength{\baselineskip}{5mm}

\noindent{\large\bf
%------------------------------------------ Title --------------------
Conformal General Relativity
%---------------------------------------------------------------------
}\vspace{4mm}

\noindent{
%-------------------------------------- Author(s) --------------------
Victor~Pervushin, Denis~Proskurin
%---------------------------------------------------------------------
}\vspace{1mm}

\noindent{\small
%------------------------------------ Address(es) --------------------
Joint Institute for Nuclear Research, 141980 Dubna, Russia
%---------------------------------------------------------------------
}\vspace{4mm}

\begin{abstract}{
The inflation-free solution of problems of the modern cosmology
(horizon, cosmic initial data, Planck era, arrow of time, singularity,
homogeneity, and so on) is considered in the conformal-invariant unified theory
given in the space with geometry of similarity where we can measure only
the conformal-invariant ratio of all quantities. Conformal General Relativity
is defined as the $SU_c(3)\times SU(2)\times U(1)$-Standard Model where the
dimensional parameter in the Higgs potential is replaced by a dilaton scalar
field described by the negative Penrose-Chernikov-Tagirov action.
Spontaneous SU(2) symmetry breaking is made on the level of
the conformal-invariant angle of the dilaton-Higgs mixing, and it allows us
to keep the structure of Einstein's theory with the equivalence principle.
We show that the lowest order of the linearized equations of motion
 solves the problems mentioned above and
describes the Cold Universe Scenario with the constant temperature T
and z-history of all masses with respect to an observable conformal time.
A new fact is the intensive cosmic creation of $W,Z$- vector bosons due
to their mass singularity. In the rigid state, this effect is determined
by the integral of motion $(m_w^2H_{\rm hubble})^{1/3}=2.7 K k_B$ that
coincides with the CMB temperature and has the meaning of the primordial
Hubble parameter. The created bosons are enough
to consider their decay as an origin of the CMB radiation and all observational
matter with the observational element abundances and the baryon asymmetry.
Recent Supernova data on the relation between the luminosity distance and
redshift (including the point $z=1.7$) do not contradict the dominance of
the rigid state of the dark matter in the Conformal Cosmology.
}
\end{abstract}

\section{Statement of the Problem}

We would like to present here the results obtained by our international
group on the construction of a unified theory of all interactions
based on the principle of relativity of all standards of measurement
~\cite{grg}-\cite{114}.
This principle can be incorporated into the unified theory
through the Weyl geometry of similarity as
a manyfold of conformal-equivalent Riemannian geometries.
To escape defects of the first Weyl version of 1918~\cite{we}, we use
the scalar-tensor conformal invariant
$(\hat g_{\mu\nu}=W^2 g_{\mu\nu})$ where $W$ is a dilaton scalar field
described by the negative Penrose-Chernikov-Tagirov (PCT) action~\cite{pct}
$$
S=-\int\limits_{ }^{ }d^4x\sqrt{-\hat g}\frac{1}{6}R(\hat g)~.
$$
This action keeps the structure of Einstein's theory
and is  a conformal analog of the Einstein-Hilbert action.
Therefore, we call this theory the Conformal General Relativity (CGR).

In contrast to Einstein's General Relativity, we can measure only a ratio
of two Einstein intervals that depends only on nine components of the
metric tensor. This means that the conformal invariance allows us to
remove only one component of the metric tensor but not the dilaton.
The best way to remove this component for the Hamiltonian
description is to use the scale-free Lichnerowicz
conformal-invariant field variables $F_{(n)}$, including the metric
$g$~\cite{L}
$$
F^L_{(n)}=||{}^{(3)}g||^{-n/6}F_{(n)}~,~~~~~
(ds)_L^2= g^L_{\mu\nu}dx^{\mu}dx^{\nu},~~~~||{}^{(3)}g^L||=1~,
$$
where $(n)$ is the conformal weight for a tensor $(n=2)$,~vector $(n=0)$,
spinor $(n=-3/2)$,~and scalar $(n=-1)$.
We show how the conformal invariance of the action, variables, and measurable
quantities allows us to solve the problems of modern cosmology
without inflation.

After the formulation of the theory and a method, we discuss
the Cold Universe Scenario where the Universe begins to form
the intensive cosmic creation of $W,Z$-vector bosons with
the constant temperature $T=(m_w^2H_{\rm hubble})^{1/3}=2.7 K$ as one of
the integrals of motion~\cite{114}.

\section{Theory}

Conformal General Relativity (CGR)
is defined as the $SU_c(3)\times SU(2)\times U(1)$-Standard Model (SM)
where the dimensional parameter in the Higgs potential
is replaced by the dilaton $W$ described by the PCT-action~\cite{114} so that
$$
{\cal L}_{\rm Higgs}=
-\lambda\left[ \left(|\Phi|\right)^{2} - y^{2}_{h}W^{2} \right]^{2}~.
$$
The conformal-invariant interactions of the dilaton and the Higgs doublet
form the effective Newton coupling in the gravitational Lagrangian
$$
\frac{|\Phi|^2-W^2}{6}R~.
$$
This coupling shows a necessity of the dilaton-Higgs mixing~\cite{vp}
$$
 W=\phi~\cosh\chi,~~~~~~|\Phi|=\phi~\sinh\chi~~~~~~~(|\Phi|^2-W^2=-\phi^2)~,
$$
so that the CGR action takes the form
$$
 {\cal L}_{CGR}= - \frac{\phi^{2}}{6}R -\partial_{\mu}\phi
\partial^{\nu}\phi + \phi^{2}\partial_{\mu}\chi\partial^{\mu}\chi+
{\cal L}_{\rm Higgs}+y_e\phi \sinh\chi \bar e e+...~,
$$
where the Higgs Lagrangian
$$
{\cal L}_{Higgs}=
 -\lambda\phi^{4}\left[\sinh^2\chi-y^{2}_{h}
\cosh^2\chi\right]^{2}
$$
describes the conformal-invariant Higgs effect of
the spontaneous SU(2)  symmetry breaking
$$
\frac{\partial {\cal L}_{\rm Higgs}}{\partial \chi}
=0~\Rightarrow
\chi_1 = 0,~~~~|\sinh \chi_2| = \frac{y_h}{\sqrt{1 - y^{2}_h}}\sim 10^{-17}~.
$$
This effect is made on the level of the mixing angle, and it
takes place even for $\lambda=0$. In this case, the trivial solution
$\chi={\rm constant}$ leads to the Higgs particle free unified
model~\cite{plb}, where any measurable ratio of masses at the same point
is a constant, and the equivalence principle is fulfilled.

The present-day value of the dilaton in the region far from heavy masses
distiguishes the scale of the Planck mass
$$
\phi (t_0,x)\simeq M_{ \rm  \bf Planck}\sqrt{\frac{3}{8\pi}}~.
$$
This fact is revealed by the energy-constrained perturbation
theory~\cite{grg}-\cite{pp}.

\section{Method}

The lowest order of the energy-constrained perturbation theory is formed
by linearization of all  equations of motion
in the class of functions with nonzero Fourier harmonics
(i.e. the "local" class of functions) in the flat conformal space-time
\be \label{cst}
ds^2_L=d\eta^2-dx_i^2~,~~~~~~~~~~~~~d\eta = N_0(t)dt~,
~~~~~~~ N_0=[g_L^{00}]^{-1/2}~.
\ee
Part of these local equations are constraints that form the projection
operators. These operators remove all superfluous degrees of freedom
of massless and massive local fields. In particular, four local
constraints as the equations for $g_{\mu\nu= 0}$  remove three longitudinal
components of gravitons and all nonzero Fourier harmonics of the
dilaton. However, the local constraints could not remove the zero
Fourier component of the dilaton
$$
 \phi^L(t,x) = \vh(t)~.
$$
The infrared interaction of the complete set of local independent
variables $\{f\}$ with this dilaton zero mode $\vh(t)$
is taken into account exactly.
The lowest  order of the considered linearized perturbation theory
is described by the Hamiltonian form of the CGR action in this
approximation
$$
S_0= \int\limits_{t_1}^{t_2}dt
 \int\limits_{V_0 }^{ } d^3x \left(\sum\limits_{f}p_f\dot f
-P_{\vh}\dot \vh -N_{0}[-\frac{P_{\vh}^2}{4}+\rho(\vh)]\right)~,
$$
where $\rho(\vh)$ is the global energy density that  generates
all the above-mentioned linear equations for independent degrees of freedom.
This energy-constrained theory contains the Friedmann-like equation for
the conformal time (\ref{cst})
\be \label{time}
\eta(\vh_0,\vh_I)=
\pm\int\limits_{\vh_I }^{\vh_0 }\frac{d\vh}{\sqrt{\rho(\vh)}}
\ee
as a consequence of the energy constraint
$$
-\frac{P_{\vh}^2}{4}+\rho_F=0
$$
and the equation  for the dilaton momentum $P_{\vh}$
$$
\frac{d\vh}{d\eta}=\frac{P_{\vh}}{2}~=\pm \sqrt{\rho(\vh)}~.
$$
The cosmic evolution of dilaton masses leads to
the redshift of energy levels of star atoms~\cite{N}
with the energy density $\rho(\vh)$ and the Hubble parameter
$H_0=\frac{\vh'}{\vh}(\eta_0)$, which gives the present day value of the dilaton
$$
 \vh(\eta_0)=
\frac{\sqrt{\rho_0}}{H_0}=\Omega_0^{1/2} M_{\rm Planck}
\sqrt{\frac{3}{8\pi}}~.
$$
Therefore, the Planck scale is distiguished as a current (present-day)
value of the dilaton,
rather than the fundamental parameter that can be shifted into the beginning
of the Universe.

The energy-constrained theory solves also the problems
of homogeneity and horizon (by the perturbation theory in the conformal
space (\ref{cst})), the positive arrow of conformal time (\ref{time})
(by the Dirac quantization
of the dilaton motion forward with a positive energy that we call
a universe, and, backward with
a negative one that we call an anti-universe), a cosmic singularity (by
nonzero boundary conditions for the dilaton in the observable Universe),
the Higgs particle, monopoles, and walls (by the consideration of the Higgs
potential-free version), the cosmic initial data (by
the diagonalization of the energy density
and the equations of motion)~\cite{pp,114}.

The energy density can be represented in the diagonal form
$$
\rho(\vh)=\sum\limits_{\varsigma }^{ }\omega_{f}(\vh,k){\hat N}_{\varsigma}
$$
(where $\omega_{f}(\vh,k)=\sqrt{k^2+y_f^2\vh^2}$ is the one-particle energy;
${\hat N}_{\varsigma} =
\frac{1}{2}(a_{\varsigma}^{+}a_{\varsigma} +
a_{\varsigma}a_{\varsigma}^{+})$ is the number of particles;
$\varsigma$ include momenta $k_i$, species
$f=h,\gamma,v,s,\chi$, spins $\sigma$),
if we introduce  "particles" as the holomorphic field variables
$$
   f(t,\vec x)=\sum\limits_{k}^{ }
\frac{C_f(\vh)\exp(ik_ix_i)}{V_0^{3/2}\sqrt{\omega_f(\vh,k)}}
\frac{1}{\sqrt{2}}\left( a_{\sigma}^+(-k,t)\epsilon_{\sigma}(-k)+
a_{\sigma}(k,t)\epsilon_{\sigma}(k)\right)~,
$$
where
$$
C_{\chi}(\vh) = \frac{\sqrt{2}}{\vh},~~~~
C_h(\vh) = \frac{\sqrt{12}}{\vh},~~~~
C_{(f=\gamma, s)}(\vh)=1,~~~~~C^{\bot}_{v}=1,~~~~
C^{||}_{v}=\frac{\omega_{v}}{y_{v}\vh}~.
$$
At the same time, the canonic diferential form in the action acquires
nondiagonal
terms as sources of cosmic creation of particles
$$
\left[\int\limits_{V_0 }d^3x \sum\limits_{f,k}p_f\dot f \right]_B
=\sum_{\varsigma = (k, f,\sigma)}
\frac{\imath}{2}( a^{+}_{\varsigma} {\dot a}_{\varsigma}
-a_{\varsigma}{\dot a}^{+}_{\varsigma} ) -
\sum_{\varsigma}
(\frac{\imath}{2}(a^{+}_{\varsigma}a^{+}_{\varsigma} -
a_{\varsigma}a_{\varsigma}))
\dot \Delta_{\varsigma}(\vh)~.
$$
The number of created particles
is calculated by the diagonalization of
equations of motion by the Bogoliubov transformation
$$
b_{\varsigma}^{+}=\cosh(r_{\varsigma})e^{-i \theta_{\varsigma}}
a_{\varsigma}^{+}
        -\imath \sinh(r_{\varsigma})e^{ i \theta_{\varsigma}}a_{\varsigma},
$$
$$
b_{\varsigma}    =\cosh(r_{\varsigma})e^{ i \theta_{\varsigma}}a_{\varsigma}
       + \imath\sinh(r_{\varsigma})e^{-i \theta_{\varsigma}}a_{\varsigma}^{+}
$$
The equations for Bogoliubov coefficients
$$
[\omega_{\varsigma} - \theta'_{\varsigma}] \sinh(2r_{\varsigma}) =
\Delta'_{\varsigma}\cos(2\theta_{\varsigma})\cosh(2r_{\varsigma}),
$$
$$
r'_{\varsigma} = - \Delta'_{\varsigma}\sin(2\theta_{\varsigma})
$$
determine the number of particles
$$
{\cal N}_{\varsigma}^{(B)}(\eta) =
{}_{sq}<0|\hat N^{(B)}_{\varsigma}|0>_{sq}-
\frac{1}{2}=\sinh^2 r_{\varsigma}(\eta)
$$
created during the time $\eta$ from
squeezed vacuum: $b_{\varsigma}|0>_{sq} = 0$ and the evolution
of the density
$$
\rho(\vh)=\vh'^{2} = \sum_{\varsigma} \omega_{\varsigma}(\vh)
{}_{sq}<0|\hat N_{\varsigma}|0>_{sq}~.
$$
The set of nondiagonal terms in SM

$
\Delta_{h}(\vh) = \ln(\vh /\vh_{I}),
$

$
\Delta^{{\bot}}_{v}(\vh) =
\frac{1}{2}\ln(\omega_{v} / \omega_{I})~,
$

$
\Delta^{||}_{v}(\vh) =\Delta_{h}(\vh) - \Delta^{\bot}_{v}(\vh)~,
$

$
\Delta_{\chi}(\vh) =\Delta_{h}(\vh) + \Delta^{\bot}_{v}(\vh)~,$\\
where $\vh_I$ and $\omega_I$ are initial data, contains the zero-mass
singularity~\cite{sf,hp} that plays an important role in the primordial
creation of longitudinal vector bosons with the properties of
the cosmic microwave background (CMB) radiation.

\section{Results}

In the limit of the Early Universe $\vh \Rightarrow 0$,
the CGR action gives the well-known
rigid state
${\rho}/{\rho_0}=\Omega_{\rm rigid}(z+1)^2$
and the primordial motion of the dilaton
$$
(\vh^2)''=0~ \Rightarrow ~\vh^2(\eta)=\vh^2_I[1+ 2 H_I \eta]=
\frac{\vh^2_0}{(1+z)^2}~,~~~
~H(z)=\frac{\vh'}{\vh}=H_0(1+z)^2~.
$$
 At the point of coincidence of the Hubble
parameter of this motion  with the mass of vector bosons
$m_v(z)\sim H(z)$, there occurs
the intensive creation of  longitudinal bosons (see Fig.1).

The temperature
of thermal equilibrium of bosons can be estimated from the restriction
for the inverse time of relaxation
$\eta^{-1}_{\rm relaxation}=\sigma_{\rm scat.}n_v(T_{eq})\geq H(z)$,
and it is  the integral of motion
$
T_{eq}\!\simeq\! [m^2_v(z)H(z)]^{1/3}\!\simeq\! (m_W^2H_0)^{1/3}=2.7 K
\!\sim\! H_I
$.
It is wonderful  that the present-day value of the boson mass and Hubble
parameter gives the Cosmic  Microwave Background temperature.

\vspace{-1cm}

\begin{center}
\begin{figure}[t]
\epsfig{figure=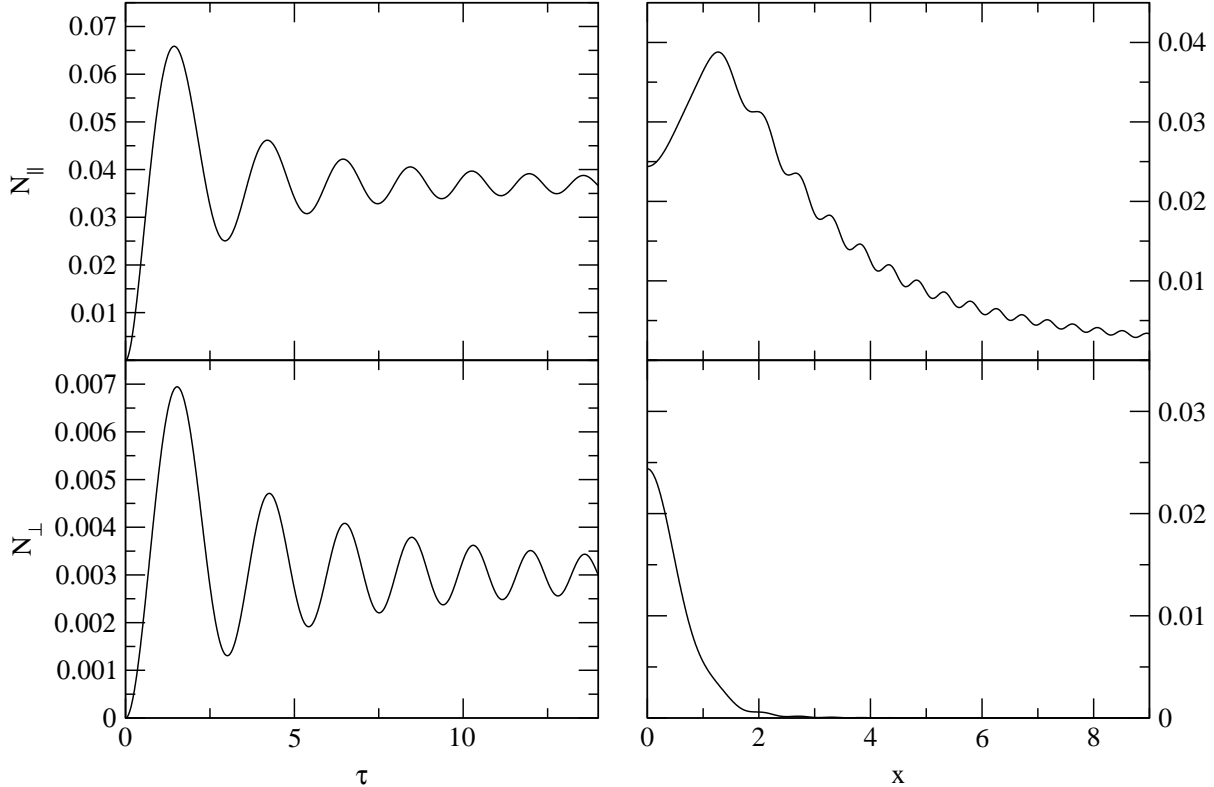,width=16cm}
%\vspace{3cm}
\caption{
Time dependence  for the dimensionless momentum
$x=k/H_I=1.25$ (left panels) and momentum dependence, at the dimensionless
lifetime
$\tau=(\eta 2H_I)=14$, (right panels) of the transverse (lower panels) and
longitudinal
(upper panels) components of the vector-boson distribution function
\cite{114}
}
\end{figure}
\end{center}

The opposite (present-day stage) limit
$\vh~ \Rightarrow~ M_{\rm Planck} \sqrt{\frac{3}{8\pi}}$
includes the dust stage ${\rho}/{\rho_0}\sim {\Omega_{\rm M}}/{(z+1)}$
with the accelerating evolution in the conformal time.
As a light ray traces a null geodesic that satisfies
the equation ${dr}/{d\eta} = 1$, the coordinate
distance as a function of the redshift z in the Conformal Cosmology (CC)
$$
H_0 r(z)=\int\limits_{1 }^{1+z }
\frac{dx}{\sqrt{\Omega_{\rm Rid.}x^6
+\Omega_{\rm Rad.}x^4+\Omega_{\rm M}x^3+\Omega_{\Lambda}}}~~~~~~~~~~~~
(\sum\limits_{I={\rm Rid., Rad., M,}\Lambda }^{}\Omega_I =1)
$$
coincides with the similar relation between coordinate
distance and redshift in Standard Cosmology (SC).

The luminosity distance $\ell$ is defined so that the apparent
luminosity of any object behaves as $1/\ell^2$. Therefore, in comparison
 with the stationary space
in SC and stationary masses in CC, a part of photons is lost. To restore the
full luminosity in both SC and CC, we should multiply the coordinate distance
by the factor $(1 + z)^2$. This factor comes from the evolution of the
angular
size of the light cone of emitted photons in SC and from the increase of the
angular size of the light cone of absorbed photons in CC.
However, in SC, we have an additional factor (1+z) due to the expansion of the
universe, since measurable distances in SC are related to measurable distances
in CC (that coincide with the coordinate ones) by the relation
\be \label{scale}
\ell = a \int \frac{dt}{a}=a r(z),
~~~~~ a = \frac{\vh}{\vh_0}= \frac{1}{1+z}~.
\ee
Thus we obtain the relations
$$
 \ell_{SC} (z) = (1 + z)^2 \ell = (1 + z)r(z)~,
$$
$$
 \ell_{CC} (z) = (1 + z)^2 r(z)~.
$$
This means that the observational data are described by different regimes in
SC and CC.
In Fig. 2, we compare the results of SC and CC for the relation
between the effective magnitude and redshift:
m(z) = 5 log [$H_0 \ell(z)$] + M where M is
a constant with
recent experimental data for distant supernovae.

\begin{center}
\begin{figure}[t]
\epsfig{figure=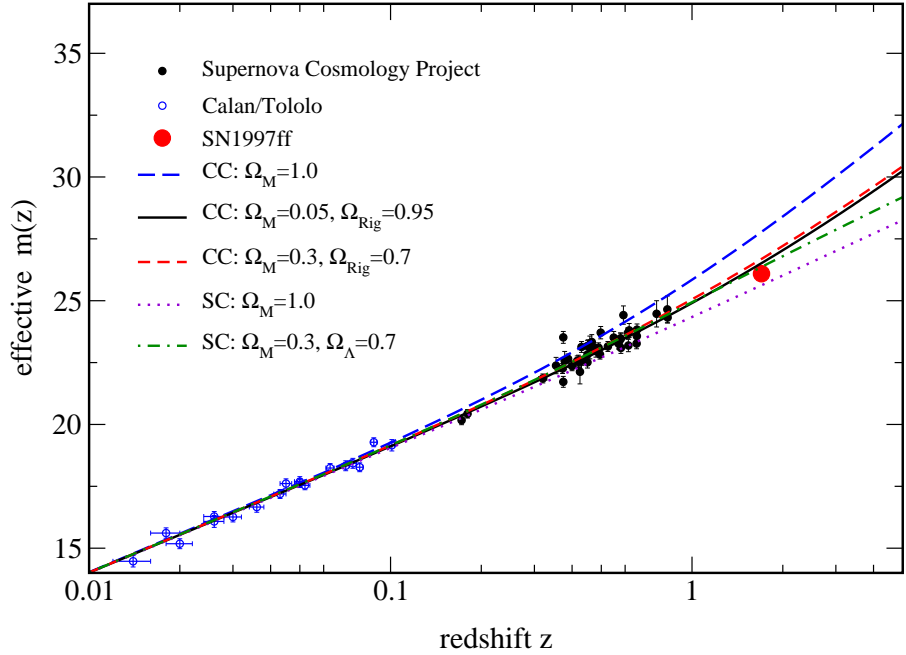,width=14cm}
%\vspace{3cm}
\caption{
m(z)- relation for a flat universe model in SC and CC. The data points
include those from 42 high-redshift Type Ia supernovae [14] and those
of recently reported farthest supernova SN1997ff [15].
An optimal fit to these data within the
SC requires the cosmological
constant $\Omega_\Lambda = 0.7$, whereas in the CC presented here
no cosmological constant is needed.
}
\end{figure}
\end{center}
There is also nonzero baryon
asymmetry due to the squeezed vacuum expectation value of the
winding number functional of the primordial
transversal bosons for their lifetime,
if we have three Sakharov conditions:
1) $CP_{\rm SM}$, 2) $H_0\not= 0$, 3)
 $\Delta L = 3 \Delta B = \Delta n_w + \Delta n_z \not= 0$

$$
\Delta n_{w,z} =\frac{{\alpha}_{w,z}}{4\pi}
\int_{0}^{\eta^{\rm w,z}_{\rm l}} d\eta
\int d^3 x~~{}_{\rm sq}<0|E^{W}_{i}B^{W}_{i}|0>{}_{\rm sq}~,
$$
$$
 {\alpha}_{w}=  \frac{4{\alpha}_{QED}}{\sin^{2}\theta_{W}},~~~
{\alpha}_{z}=\frac{{\alpha}_{QED}}{\sin^{2}\theta_{W}\cos^{2}\theta_{W}},
\eta_l^w=15 H_I^{-1},~~
\eta_l^z=30 H_I^{-1}~.
$$

Thus, we can propose the Cold Universe Scenario with the z-history of
masses and the invariant temperature $\sim 2.7 K$.
The Universe begins from the rigid primordial
motion of the modulus of the dilaton-Higgs mixing. This motion creates
vector bosons, the decay of which forms all matter in the Universe
with the constant temperature that coincides with the primordial
Hubble parameter. There are arguments in  favor of that the Cold
Universe Scenario reproduces all results of the Hot one on the
primordial element abundances in the radiation stage,
since we have in CC the same (square root) dependence of the scale factor
(\ref{scale}) on the observable time in the rigid stage
and the same argument for the Boltzmann factors.
In contrast to SC, this rigid stage dynamics of the chemical evolution
in CC does not contradict recent Supernova data.

We would like to thank the Organizing Committee,
especially Prof. M.Yu. Khlopov, for hospitality.
We are indebted to Prof. D. Blaschke and Prof. S. Vinitsky
for stimulating discussions and collaboration.
The authors are also grateful to D. Behnke and A. Gusev
for their help with the numerical work.

D.P. thanks RFBR (grant 00-02-81023Bel\_a) for support.

\end{document}